\documentclass[sigconf, nonacm]{acmart}
\newcommand\vldbdoi{XX.XX/XXX.XX}

\newcommand\vldbavailabilityurl{URL_TO_YOUR_ARTIFACTS}
 
\newcommand{\eat}[1]{}

\usepackage{latexsym}
\usepackage{amsfonts}
\usepackage{amsmath}

\usepackage{pifont}

\usepackage{xcolor}
\usepackage{colortbl}
\usepackage{epsfig}
\usepackage{xspace}
\usepackage{graphicx}
\usepackage{subfigure}
\usepackage{paralist}
\usepackage{enumerate}
\usepackage[color,matrix,arrow,all]{xy}
\usepackage{comment}
\usepackage{booktabs}
\usepackage{balance}
\usepackage{stmaryrd}
\usepackage{pifont}
\usepackage{hhline}
\usepackage{listings}
\usepackage{array}
\usepackage{float}
\usepackage[flushleft]{threeparttable}

\usepackage{mathrsfs}
\usepackage{makecell}

\usepackage{wrapfig}

\usepackage{epsfig}
\usepackage{multirow}
\usepackage{url}
\usepackage{algorithmic}
\usepackage{graphicx}
\usepackage{textcomp}
\usepackage{xcolor}

\usepackage{makecell}
\usepackage{bbm}

\usepackage{color}
\usepackage{framed}

\definecolor{shadecolor}{rgb}{0.92,0.92,0.92}

\definecolor{inputcolor}{RGB}{255,139,35}
\definecolor{outputcolor}{RGB}{120,212,252}
\definecolor{embedcolor}{RGB}{254,127,156}
\definecolor{maskcolor}{RGB}{122,128,255}
\definecolor{ecolor}{RGB}{58,149,54}

\usepackage{tikz}
\usetikzlibrary{shapes,snakes}
\usetikzlibrary{calc}

\usepackage[export]{adjustbox}

\newcommand{\at}[1]{\protect\ensuremath{\mathsf{#1}}}

\newcommand{\be}{\begin{enumerate}}
	\newcommand{\ee}{\end{enumerate}}
\newcommand{\beqn}{\begin{eqnarray*}}
	\newcommand{\eeqn}{\end{eqnarray*}}

\newcommand{\stitle}[1]{\vspace{1mm}\noindent{\bf #1}}
\newcommand{\etitle}[1]{\vspace{1mm}\noindent{\underline{\em #1}}}
\newcommand{\ie}{{\em i.e.,}\xspace}
\newcommand{\eg}{{\em e.g.,}\xspace}

\newcommand{\bi}{\begin{itemize}}
\newcommand{\ei}{\end{itemize}}

\makeatletter
\newcommand\figcaption{\def\@captype{figure}\caption}
\newcommand\tabcaption{\def\@captype{table}\caption}
\makeatother

\tikzstyle{mybox} = [draw=black, fill=black!5, thick,
rectangle, rounded corners, inner sep=0pt, inner ysep=6pt]
\tikzstyle{fancytitle} =[fill=black, text=white]

\makeatletter
\newif\if@restonecol
\makeatother

\usepackage[lined,boxed,vlined,ruled]{algorithm2e}
\usepackage{hyperref}
\newcommand{\sys}{\textsc{ChatPipe}\xspace}
\newcommand{\hci}{Human-ChatGPT\xspace}

\newcommand{\term}[1]{{\tt #1}}

\begin{document}

\title{\sys: Orchestrating Data Preparation Program by Optimizing Human-ChatGPT Interactions}
\author{Sibei Chen} 
\affiliation{%
  \institution{Renmin University, China}
}
\email{sibei@ruc.edu.cn}

\author{Hanbing Liu} 
\affiliation{%
  \institution{Renmin University, China}
}
\email{liuhanbing@ruc.edu.cn}

\author{Weiting Jin} 
\affiliation{%
  \institution{Renmin University, China}
}
\email{waiting_jin@ruc.edu.cn}

\author{Xiangyu Sun} 
\affiliation{%
  \institution{Renmin University, China}
}
\email{sunxiangyu0906@ruc.edu.cn}

\author{Xiaoyao Feng} 
\affiliation{%
  \institution{Renmin University, China}
}
\email{fengxiaoyao668@ruc.edu.cn}

\author{Ju Fan} 
\affiliation{%
  \institution{Renmin University, China}
}
\email{fanj@ruc.edu.cn}

\author{Xiaoyong Du} 
\affiliation{%
  \institution{Renmin University, China}
}
\email{duyong@ruc.edu.cn}

\author{Nan Tang} 
\affiliation{%
  \institution{QCRI}
}
\email{ntang@hbku.edu.qa}

\begin{abstract}

Orchestrating a high-quality data preparation program is essential for successful machine learning (ML), but it is known to be time and effort consuming. Despite the impressive capabilities of large language models like ChatGPT in generating programs by interacting with users through natural language prompts, there are still limitations. Specifically, a user must provide specific prompts to iteratively guide ChatGPT in improving data preparation programs, which requires a certain level of expertise in programming, the dataset used and the ML task. Moreover, once a program has been generated, it is non-trivial to revisit a previous version or make changes to the program without starting the process over again.

In this paper, we present \sys, a novel system designed to facilitate seamless interaction between users and ChatGPT. \sys provides users with effective recommendation on next data preparation operations, and guides ChatGPT to generate program for the operations. Also, \sys enables users to easily roll back to previous versions of the program, which facilitates more efficient experimentation and testing.
We have developed a web application for \sys and prepared several real-world ML tasks from Kaggle. These tasks can showcase the capabilities of \sys and enable VLDB attendees to easily experiment with our novel features to rapidly orchestrate a high-quality data preparation program.

\end{abstract}
\maketitle




\section{Introduction}
\label{sec:intro}
As an indispensable step in data analysis, data preparation (or \emph{data prep} for short) transforms raw data into a format that is ready to use for machine learning (ML) or data science tasks. A typical data prep process involves a series of \emph{operations}, such as handling missing or invalid data, normalizing data, and feature engineering, where data moves from one operation to subsequent operations. In this paper, we call such series of operations as \emph{data prep program}.


In practice, domain experts, like physicians, may find it challenging to orchestrate data prep program due to the need for programming skills and data prep experiences.
Recently, large language models, especially ChatGPT, have emerged as a promising tool to assist domain experts in orchestrating data prep program. Different from existing data prep tools, ChatGPT can directly generate a data prep program by simply ``chatting'' with users.





However, it is non-trivial for users to provide specific \emph{prompts} to guide ChatGPT for improving the data prep program. The main reason is that users may not know which operations are the most helpful given a specific dataset and a particular task. Moreover, it is also hard for users to revisit previous program versions, if the current program produces inferior performance.
Therefore, it is highly desirable to equip users with an \emph{assistant} that can provide effective guidance by suggesting ChatGPT the operations to improve the program, and easily roll back to any previous versions to facilitate experimentation and testing.

\begin{figure}
	\centering
	\includegraphics[width=0.5\textwidth]{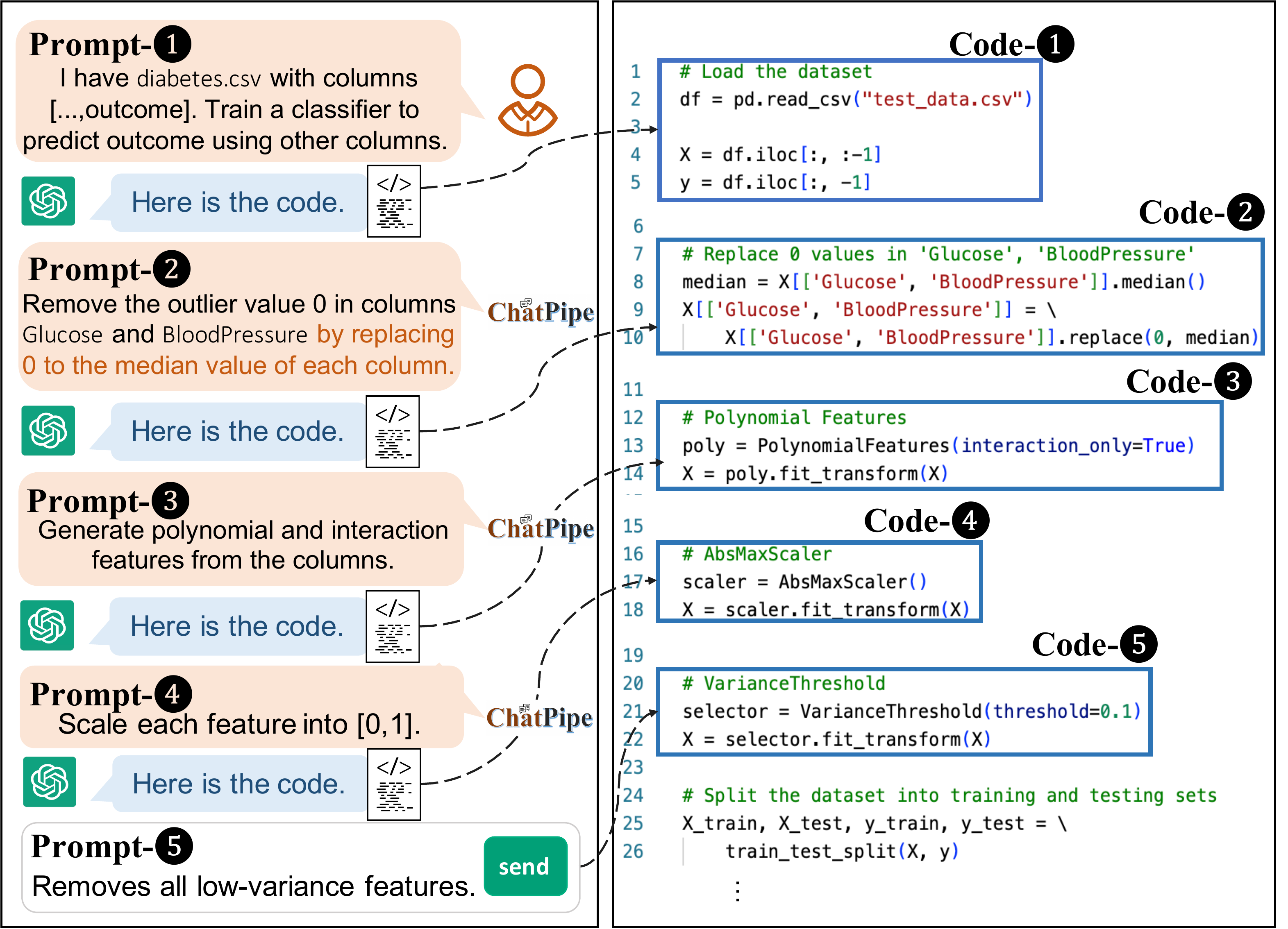}
	\vspace{-2em}
	\caption{An overview of \sys, which provides guidance by suggesting operations to improve \hci interactions for data prep program orchestration.}
	\label{fig:prototype}
	\vspace{-2em}
\end{figure}

\stitle{Our proposal.}
We propose \sys, a tool to provide effective \emph{guidance} by suggesting operations to improve \hci interactions for data prep program orchestration. As illustrated in Figure~\ref{fig:prototype}, \sys enables seamless communication between a user and ChatGPT, and generates a data prep program as follows.  

(1) Initially, the user uploads a dataset (\eg~\term{diabetes}.\term{csv}) and provides a prompt (\ie \textbf{Prompt-\ding{182}}) to ChatGPT. Then, ChatGPT generates an initial code block, namely \textbf{Code-\ding{182}}, which loads the dataset and prepares features and labels.

(2) \sys analyzes the dataset and detects outlier values, such as $0$, in columns \term{Glucose} and \term{BloodPressure}, and thus recommends an \emph{outlier removal} operation. Moreover, \sys enables the user to update the prompt to inject domain knowledge, \eg replacing all $0$s with median of each column. Guided by the prompt (\textbf{Prompt-\ding{183}}), ChatGPT generates a code block \textbf{Code-\ding{183}}.

(3) By interacting with the user, \sys recommends a series of operations \term{PolynomialFeatures} (\textbf{Prompt-\ding{184}}), \term{MaxAbsScaler} (\textbf{Prompt-\ding{185}}), and \term{VarianceThreshold} (\textbf{Prompt-\ding{186}}), which are tailored for feature engineering. We can see that these suggested operations are quite reasonable and coherent, \ie first generating interaction features from columns, then scaling the generated features, and finally removing low-variance features.

In such a way, \sys can speed up the time-consuming ``trial-and-error'' process of data prep by guiding ChatGPT to generate effective data prep program via \hci interactions.

\stitle{Architecture of \sys.}
To support the aforementioned features, \sys consists of three key modules.
(1) \emph{Operation Recommendation:} Given a current program, \sys recommends the most effective operations to improve the overall performance of the ML task. This module is powered by the techniques proposed in our research paper~\cite{haipipe}.
(2) \emph{Program Generation:} Given the recommended operations, \sys interacts with ChatGPT to generate an error-free data prep program that effectively implements the operations.
(3) \emph{Version Management:} The \hci interactions would generate multiple program versions, including produced datasets and corresponding prompts. To enable users to roll back to any previous versions, \sys visualizes the program versions, and develops techniques to speed-up program execution.

\stitle{Demonstration scenarios.} We build \sys as a web application and demonstrate it by orchestrating data prep programs for real ML tasks from Kaggle, the most popular data science website. We prepare a collection of ML tasks, each of which contains a dataset, such as \term{diabetes}.\term{csv} in Figure~\ref{fig:prototype}. We allow the VLDB attendees to choose ML tasks, and then leverage \sys to interact with ChatGPT for generating effective data prep programs. During the process, we illustrate the novel features of \sys, including operation recommendation, program generation, and version management. A demonstration video can be found on YouTube\footnote{\url{https://youtu.be/bHwEXaA9y6Y}}.

To summarize, we make the following contributions.
(1) We develop \sys, a novel tool to optimize Human-ChatGPT interactions for data prep program orchestration.
(2) \sys can provide effective guidance to ChatGPT by suggesting operations to generate program, and enable users to jump over different program versions.
(3) We deploy \sys as a web app with user-friendly interface and demonstrate its utility on real data prep scenarios.

\section{System Overview}

\begin{figure}
	\centering
	\includegraphics[width=\columnwidth]{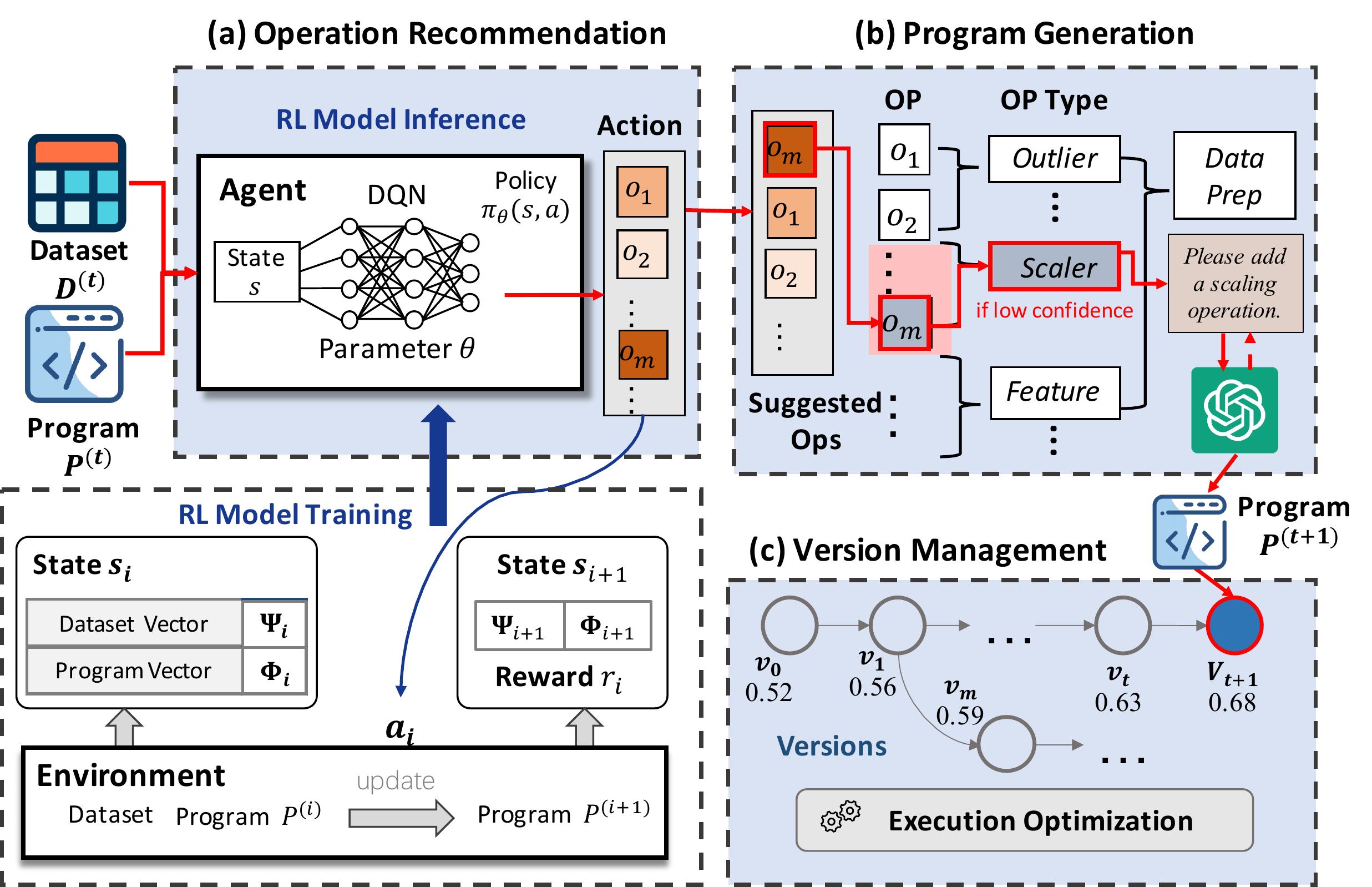}
	\vspace{-2em}
	\caption{Architecture of \sys, which consists of three key modules, (a) Operation Recommendation, (b) Program Generation, and (c) Version Management.}
	\label{fig:framework}
		\vspace{-1em}
\end{figure}

The architecture of \sys is shown in Figure~\ref{fig:framework}. Given an ML task, such as training a classifier for diabetes prediction based on a dataset, \sys is designed to optimize \hci interactions for orchestrating data prep programs. To achieve this goal, \sys consists of three key modules.
\emph{Operation Recommendation} provides users with effective recommendations on next data prep operations, such as dealing with missing values and feature engineering, based on the current program and dataset (Section~\ref{sec:operation_selection}).
\emph{Program Generation} interacts with ChatGPT to generate an error-free data prep program that effectively implements the recommended operations (Section~\ref{sec:prompt_gen}).
\emph{Version Management} visualizes various program versions generated by the \hci interactions, and enables users to rapidly roll back to previous versions for more efficient experimentation and testing (Section~\ref{sec:pipe_exe_opt}).

\subsection{Operation Recommendation} \label{sec:operation_selection}
The key challenge of \emph{Operation Recommendation} is two-fold. First, there are a variety of data prep operation types (\eg \emph{Scaling} and \emph{Discretization}), and even for one type, there are many possible operations (\eg \term{MinMaxScaler} and \term{StandardScaler}). Second, there could be many possible ways of recommending operations, which are highly dataset-specific and task-specific.
To address the problem, we introduce a \emph{reinforcement learning (RL)} based approach that learns how to recommend operations through offline training. For inference, given a dataset and the current program, we use the trained model to suggest the most appropriate operations.
More details of our approach can be found in our research paper~\cite{haipipe}.

\stitle{Reinforcement Learning Framework.}
We formulate the operation recommendation problem as a sequential decision process by an \emph{agent}. Based on a \emph{policy}, the agent takes the current program $P^{(i)}$ and dataset $D^{(i)}$ as \emph{state}, and performs an \emph{action}, \ie selecting an operation from a pre-defined search space. Then, the agent obtains \emph{rewards} from an \emph{environement}, and updates its policy accordingly.

\etitle{State:} We represent the state based on the current program $P^{(i)}$ and the produced dataset $D^{(i)}$. Specifically, we use statistical information as dataset features, and leverage a public pre-trained code representation model UniXcoder~\cite{unixcoder} to extract program features. Then, we concatenate the features to represent the state.

\etitle{Action:} We design action as choosing an operation $o_i$ from a pre-defined operation set $O$, which transforms program $P^{(i)}$ to $P^{(i+1)}$.

\etitle{Reward:} The reward is used as a signal of whether the actions performed are beneficial. We execute program $P^{(i+1)}$ to evaluate its performance, \eg F1-score of the trained classifier.

%

\stitle{Deep Q-Network (DQN).}
We adopt a Deep Q-Network (DQN) framework to optimize the \emph{policy} function, as shown in Figure~\ref{fig:framework}. Specifically, following the typical strategy in DQN, we adopt a neural network to approximate the value function in DQN, which takes state as input and produces a value for each action (\ie operation). The neural network contains multiple fully connected layers with \term{LeakeyRelu} as activation function, and an output \term{tanh} layer.

\stitle{DQN Training and Inference.}
Figure~\ref{fig:framework} illustrates the processes of DQN training and inference. 
Our DQN training algorithm uses an off-policy strategy that learns an $\epsilon$-greedy policy in multiple episodes. Specifically, in the $i$-th episode, it first computes current state $s_i$, and then selects the greedy action that maximizes the value function with a probability $1-\epsilon$, and a random action with probability $\epsilon$. Given the selected action $a_i$ (\ie operation), the algorithm updates the program from $P^{(i)}$ to $P^{(i+1)}$, and executes $P^{(i+1)}$ to obtain reward $r_i$. Then, the parameters of DQN can be updated using stochastic gradient descent.

For inference, given a new program $P^{(t)}$ and dataset $D^{(t)}$, we extract their features to represent the state, and then utilize the trained model to compute values for the operations in our candidate set. Finally, we maintain the operations as a ranked list, and feed them to the \emph{Program Generation} module introduced below.
\vspace{-1em}
\subsection{Program Generation} \label{sec:prompt_gen}

\emph{Program Generation} interacts with ChatGPT to generate data prep programs through natural language \emph{prompts}.  The challenge here is how to guide ChatGPT to generate an error-free data prep program that effectively implements the operations. To address the challenge, we develop three components, \emph{Operation Augmentation}, \emph{Program Refinement}, and \emph{Program Checking}, as described below.

\stitle{Operation Augmentation.}
A straightforward method is to directly ``translate'' the best operation returned by \emph{Operation Recommendation}. For example, for the \term{StandardScaler} operation, one prompt could be ``\emph{Standardize features by removing the mean and scaling to unit variance}''. However, as the operation is from a constrained set of candidate operations (see Section~\ref{sec:operation_selection}), in some cases, the improvement in the operation may not be significant. 

Fortunately, we have an interesting observation that ChatGPT may be helpful in these cases. Specifically, when we guide ChatGPT with a \emph{coarse-grained} prompt, \eg ``\emph{Generate interaction features}'', it may generate a customized program, such as \term{df[`\textit{AB}'] = df[`\textit{A}'] * df[`\textit{B}']}, instead of simply using predefined operations. 
This inspires us to organize operations into a two-level hierarchy, where the bottom layer is fine-grained operations in our pre-defined operations set (\eg~\term{sklearn.preprocessing.MinMaxScaler}), and the upper layer is the type of operators (\eg~\term{Scaler}). Based on the two-level hierarchy, we generate the prompts as follows.
When the \emph{confidence} of the recommended operations is high, \eg the output score of DQN is greater than a threshold, we select the fine-grained operations directly. Otherwise, we select the best operation type, which has the highest average score of its fine-grained operations, to generate a coarse-grained prompt.

\stitle{Program Refinement.} This component asks the users to refine the generated program by injecting their domain knowledge or data prep experiences. 
For example, a user can utilize domain knowledge to determine how to perform discretization on BMI, \eg ``The BMI should be cut into $[0, 18.5, 25, 30, 100]$ with [`underweight', `normal', `overweight', `obese']''. Moreover, the users can also customize parameters of the recommended operations, either by refining the prompts or by updating the program.


\stitle{Program Checking.} This component checks whether the generated program has run-time errors, and fixes the errors if there is any. To this end, before sending the program to users, \sys first runs the generated program in a local run-time environment. If some errors occur, \eg ``\textit{NameError: name 'PolynomialFeatures' is not defined.}", \sys sends the error log to ChatGPT, and ChatGPT can fix the error to provide an updated program. Consider the above example again. ChatGPT will add one line in the head of program: ``\textit{from sklearn.preprocessing import PolynomialFeatures}". It iteratively runs this process until the program is error-free.

\begin{figure*}
	\centering
	\includegraphics[width=\textwidth]{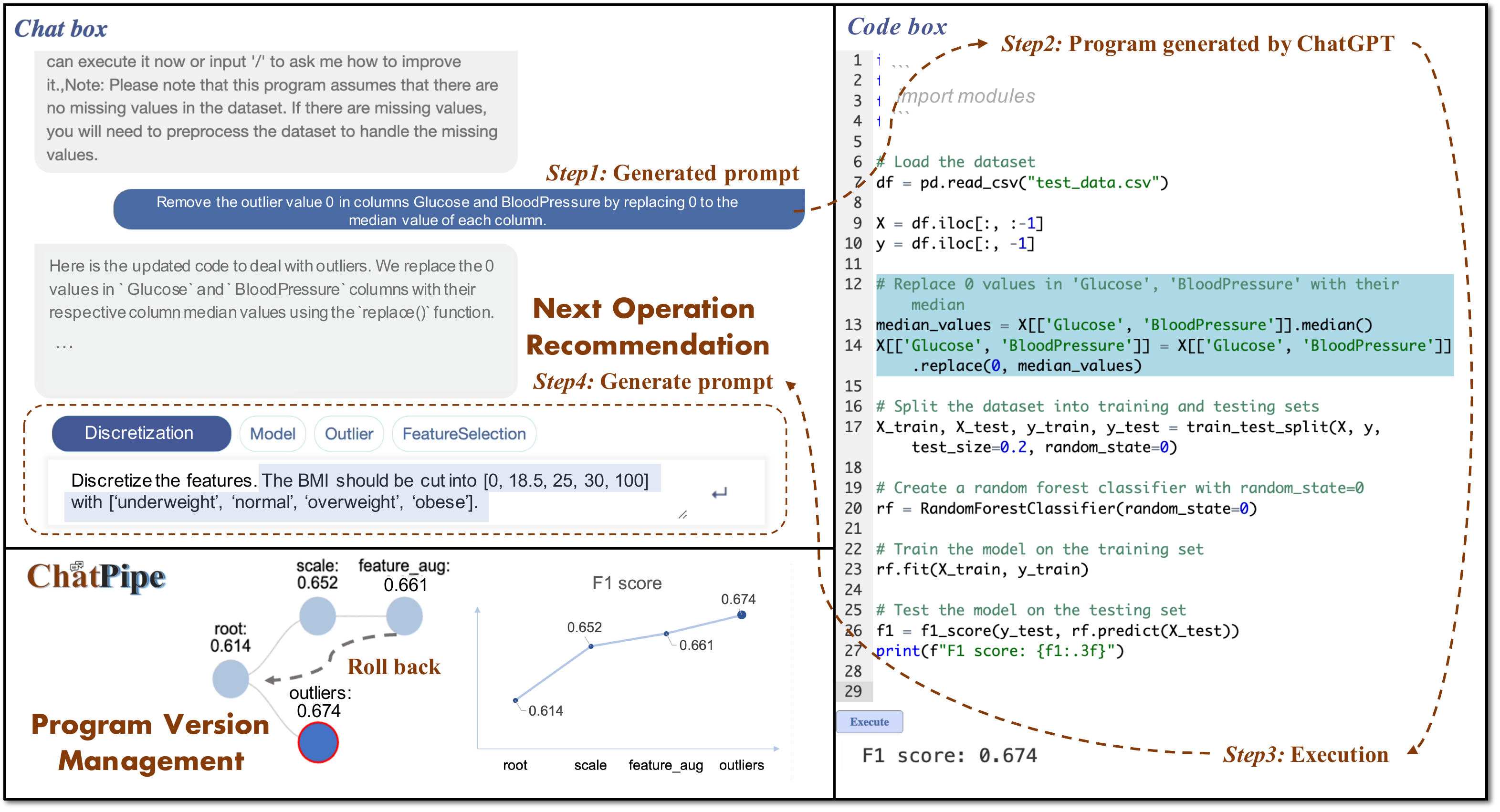}
	\vspace{-2em}
	\caption{Demonstration scenarios of \sys.}
	\label{fig:demo}
		\vspace{-1em}
\end{figure*}

\subsection{Version Management} \label{sec:pipe_exe_opt}
\hci interactions would generate multiple versions of the program, including the produced datasets and the corresponding prompts. To enable users to roll back to any previous versions, \emph{Version Management} is designed to visualize the program versions and support fast version switching. This module has two components: (1) Program Versioning, which maintains versions between programs, and (2) Data Caching, which accelerates the execution by caching intermediate data variables across versions. 

\stitle{Program Versioning.} 
We use a relational database to store version-related information for each ML task, where each version contains the corresponding program, executing results, and the prompts. We also maintain the relationship between versions, which can be used to visualize the versions as a tree structure (see Figure~\ref{fig:framework}). 
Moreover, users can add new versions by interacting with ChatGPT, delete useless versions, and can choose any two versions to compare theirs differences in program, data and prompts. 

\stitle{Data Caching across Versions.}
There could be slight differences among program versions, as the versions may have a large proportion of code-blocks in common. Thus, it is natural to cache some intermediate variables, which are likely to be reused across versions, so as to enable fast version switching.

We have studied two technical challenges in data caching across versions. Firstly, some studies have shown that it is unrealistic to store all the intermediate variables~\cite{CO}. Thus, it is necessary to select the most ``beneficial'' variables for caching.
To address the challenge, we formulate the caching decision as a binary classification problem, which, given a data variable, estimates the probability that the variable is beneficial to cache. We devise a deep neural network to predict the caching probability, taking as features the variable's storage size, calculation time, previous operations and the accuracy of the downstream ML task.
The second challenge is that we should determine when a cached variable can be reused. For example, if the code-block that computes the variable is changed, the cached data is then outdated, and thus cannot be reused.
To address this, we use Abstract Syntax Tree (AST) to analyze the trace that each variable is computed, and use a optimized maximum flow algorithm~\cite{CO} to judge whether a variable should be reused or recalculated,

\section{Demonstration Scenarios}


For our upcoming demonstration, we have curated a selection of $62$ machine learning tasks from over $10$ diverse domains, including education, medicine and finance, sourced from Kaggle. Attendees at VLDB are welcome to choose a dataset or domain that they are most comfortable working with.

Next, we will provide a concrete example of how to use \sys, highlighting two key features. Firstly, \sys can assist users in their interaction with ChatGPT by suggesting the next operations. Secondly, \sys can manage user history, allowing them to roll back to a previous version and compare the differences.


\stitle{Machine learning tasks.}
Our sample machine learning task is to predict whether a patient has diabetes~\cite{diabetes}. It contains $769$ tuples and eight numeric diagnostic measurements as features. We use $\at{sklearn.model\_selection.train\_test\_split}$ to split the train set and the test set, with $\at{random\_state = 0}$ by default.


\stitle{Setup.} 
The user uploads the diabetes dataset to \sys and specifies the column name or index to be used for classification. She can then generate the initial version of the code by either writing it by hand in the Code Box or uploading an existing code. Alternatively, she can use \sys to generate the code for them. Once the code is generated, it can be executed and saved. Subsequently, the user iteratively interacts with \sys to improve the F1 score for the classification task. 

Figure~\ref{fig:demo} illustrates an example where the user has interacted with \sys four times and is about to initiate their fifth interaction, as described as follows.


\etitle{(1) Next Operation Recommendation.}
In Figure~\ref{fig:demo}, the user confirms the operation (or prompt) recommended by \sys, ``\emph{Remove the outlier value 0 in columns Glucose $\ldots$}'', which will be sent to ChatGPT to update the code, as shown in the Code Box of Figure~\ref{fig:demo}. The light-blue section highlights the added outlier handling operation. 

Afterwards, the user executes the new code and observes an improvement in F1 (0.674 vs. 0.614). Then, the user enters the ``/'' symbol, which triggers \sys to display several recommended operations (\eg text bubbles above the input box). The operations are sorted and displayed by \sys. After selecting an operation (in this case, ``\term{Discretization}''), the corresponding prompt (``Discretize the features'') is automatically filled into the input box. 

The user can also apply domain knowledge and add an additional prompt, such as ``\emph{The BMI should be cut into $[0, 18.5, 25, 30, 100]$ with [`underweight', `normal', `overweight', `obese']}''.

\etitle{(2) Program Version Management.}
\sys stores all updated and executed programs, which can be viewed in the code editor. And the executed results (\eg F1 score) of different programs can be viewed as a statistical line chart.
\sys manages these information as versions, which are organized into a version tree. Users can change the current version by clicking on the relevant node. In Figure~\ref{fig:demo}, the user starts with the initialization version of the code root, generated automatically by \sys. She then adds a scaling operation via a prompt, followed by a feature augmentation operation. During the third interaction, she decides to explore different options and clicks on the root node to return to the original version. Then, she selects the outlier handling operation further exploration.




\bibliographystyle{ACM-Reference-Format}
\bibliography{chatpipe}


\begin{thebibliography}{4}


\ifx \showCODEN    \undefined \def \showCODEN     #1{\unskip}     \fi
\ifx \showDOI      \undefined \def \showDOI       #1{#1}\fi
\ifx \showISBNx    \undefined \def \showISBNx     #1{\unskip}     \fi
\ifx \showISBNxiii \undefined \def \showISBNxiii  #1{\unskip}     \fi
\ifx \showISSN     \undefined \def \showISSN      #1{\unskip}     \fi
\ifx \showLCCN     \undefined \def \showLCCN      #1{\unskip}     \fi
\ifx \shownote     \undefined \def \shownote      #1{#1}          \fi
\ifx \showarticletitle \undefined \def \showarticletitle #1{#1}   \fi
\ifx \showURL      \undefined \def \showURL       {\relax}        \fi
\providecommand\bibfield[2]{#2}
\providecommand\bibinfo[2]{#2}
\providecommand\natexlab[1]{#1}
\providecommand\showeprint[2][]{arXiv:#2}

\bibitem[\protect\citeauthoryear{Chen, Tang, Fan, Yan, Chai, Li, and Du}{Chen
  et~al\mbox{.}}{pear}]%
        {haipipe}
\bibfield{author}{\bibinfo{person}{Sibei Chen}, \bibinfo{person}{Nan Tang},
  \bibinfo{person}{Ju Fan}, \bibinfo{person}{Xuemi Yan},
  \bibinfo{person}{Chengliang Chai}, \bibinfo{person}{Guoliang Li}, {and}
  \bibinfo{person}{Xiaoyong Du}.} \bibinfo{year}{2023 (to appear)}\natexlab{}.
\newblock \showarticletitle{HybridPipe: Combining Human-generated and
  Machine-generated Pipelines for Data Preparation}. In
  \bibinfo{booktitle}{\emph{{SIGMOD}}}.
\newblock


\bibitem[\protect\citeauthoryear{Dataset}{Dataset}{[n.d.]}]%
        {diabetes}
\bibfield{author}{\bibinfo{person}{Kaggle~Diabetes Dataset}.}
  \bibinfo{year}{[n.d.]}\natexlab{}.
\newblock
  \bibinfo{title}{\url{https://www.kaggle.com/datasets/akshaydattatraykhare/diabetes-dataset}}.
\newblock
\newblock


\bibitem[\protect\citeauthoryear{Derakhshan, Rezaei~Mahdiraji, Abedjan, Rabl,
  and Markl}{Derakhshan et~al\mbox{.}}{2020}]%
        {CO}
\bibfield{author}{\bibinfo{person}{Behrouz Derakhshan},
  \bibinfo{person}{Alireza Rezaei~Mahdiraji}, \bibinfo{person}{Ziawasch
  Abedjan}, \bibinfo{person}{Tilmann Rabl}, {and} \bibinfo{person}{Volker
  Markl}.} \bibinfo{year}{2020}\natexlab{}.
\newblock \showarticletitle{Optimizing machine learning workloads in
  collaborative environments}. In \bibinfo{booktitle}{\emph{{SIGMOD}}}.
  \bibinfo{pages}{1701--1716}.
\newblock


\bibitem[\protect\citeauthoryear{Guo, Lu, Duan, Wang, Zhou, and Yin}{Guo
  et~al\mbox{.}}{2022}]%
        {unixcoder}
\bibfield{author}{\bibinfo{person}{Daya Guo}, \bibinfo{person}{Shuai Lu},
  \bibinfo{person}{Nan Duan}, \bibinfo{person}{Yanlin Wang},
  \bibinfo{person}{Ming Zhou}, {and} \bibinfo{person}{Jian Yin}.}
  \bibinfo{year}{2022}\natexlab{}.
\newblock \showarticletitle{Unixcoder: Unified cross-modal pre-training for
  code representation}.
\newblock \bibinfo{journal}{\emph{arXiv:2203.03850}} (\bibinfo{year}{2022}).
\newblock


\end{thebibliography}

\end{document}